\documentstyle[onecolumn,colordvi]{mn}
\newcommand{\mincir}{\raise 
-2.truept\hbox{\rlap{\hbox{$\sim$}}\raise5.truept
\hbox{$<$}\ }}
\newcommand{\magcir}{\raise 
-2.truept\hbox{\rlap{\hbox{$\sim$}}\raise5.truept
\hbox{$>$}\ }}
\newcommand{\minmag}{\raise-2.truept\hbox{\rlap{\hbox{$<$}}\raise 
6.truept\hbox
{$>$}\ }}
\newcommand{\be}{\begin{equation}}
\newcommand{\ee}{\end{equation}}
\newcommand{\ba}{\begin{eqnarray}}
\newcommand{\ea}{\end{eqnarray}}
\newcommand{\brr}{\begin{array}}
\newcommand{\err}{\end{array}}
\newcommand{\bc}{\begin{center}}
\newcommand{\ec}{\end{center}}
\newcommand{\br}{\mbox{\bf r}}
\newcommand{\bv}{\mbox{\bf v}}

\newcommand{\bg}{\mbox{\bf g}}

\def\ifm#1{\relax\ifmmode#1\else$\mathsurround=0pt #1$\fi}
\def\kms{\ifmmode\,{\rm km}\,{\rm s}^{-1}\else km$\,$s$^{-1}$~\fi}
\def\hmpc{\ifmmode\,{\it h }^{-1}\,{\rm Mpc }\else $h^{-1}\,$Mpc\,\fi}

\def\fig #1, #2, #3 {
  \smallskip
  \centerline{\psfig{figure=#1,height=#2 in,width=#3 in}} }
\def\capt{\small \baselineskip 12pt }

\def\\{\hfill\break}

\def\etal{{\it et al.\ }}

\def\ifm#1{\relax\ifmmode#1\else$\mathsurround=0pt #1$\fi}
\def\kmsn{\ifmmode\,{\rm km}\,{\rm s}^{-1}\else km$\,$s$^{-1}$\fi} 
\def\kms{\ifmmode\,{\rm km}\,{\rm s}^{-1}\else km$\,$s$^{-1}$~\fi} 

\def\ltsima{$\; \buildrel < \over \sim \;$}
\def\lsim{\lower.5ex\hbox{\ltsima}}
\def\gtsima{$\; \buildrel > \over \sim \;$}
\def\gsim{\lower.5ex\hbox{\gtsima}}
 
\def\pmb#1{\setbox0=\hbox{#1}%
 \kern-.025em\copy0\kern-\wd0
 \kern.05em\copy0\kern-\wd0
 \kern-.025em\raise.0433em\box0}

\def\v0{\pmb{$0$}}

\title [SFI vs. PSC$z$]{Comparing the SFI Peculiar Velocities
with the PSC$z$ Gravity Field: a VELMOD analysis}
\author[Branchini \etal]
{E. Branchini$^{1,2}$, W. Freudling$^{3,4}$, L.N. Da Costa$^{4}$, C.S.
Frenk$^5$\\
\vspace{-1mm}\\ 
{\LARGE   R. Giovanelli$^{6}$, M.P. Haynes$^{6}$, J.J. Salzer$^{7}$,
G. Wegner$^{8}$, I. Zehavi$^{9}$}\\ 
\vspace{-1mm}\\ 
$^1$ Kapteyn Institute, University of Groningen, Landleven 12, P.O. Box
800, 9700, Groningen, The Netherlands.\\
$^2$ Dipartimento di Fisica dell'Universit\'a degli Studi ``Roma TRE'',
Via della
Vasca Navale 84, I-00146, Roma, Italy.\\
$^3$ Space Telescope  -- European Coordination Facility,
Karl-Schwarschild Strasse 2, 85748, Garching, Germany.\\
$^4$ European Southern Observatory, 
Karl-Schwarschild Strasse 2, 85748, Garching, Germany.\\
$^5$ Department of Physics, University of Durham, South Rd., DH1 3LE,
Durham, U.K. \\
$^6$ Center for Radiophysics and Space Research and National Astronomy and
Ionosphere Center. \\ 
Cornell University, Ithaca, NY 14853, USA. \\
$^7$ Astronomy Department, Wesleyan University, Middletown, CT 06457,
USA. \\
$^8$ Department of Physics and Astronomy, Darmouth College, Hanover, NH
03755, USA.\\
$^9$ NASA/Fermilab Astrophysiscs Group, Fermi National Accelerator Laboratory, 
Batavia, IL 60510-0500, USA.\\
}

\begin{document}

\maketitle
\begin{abstract}

We compare the peculiar velocities derived from the $I$-band Tully-Fisher
(TF) relation for 989 field spiral galaxies in the SFI catalogue with the
predicted velocity field derived from the IRAS PSC$z$ galaxy redshift
survey. We assume linear gravitational instability theory and apply the
maximum likelihood technique, VELMOD (Willick \etal 1997b), to SFI galaxies
within cz$<6000$ \kmsn. The resulting calibration of the TF relation is
consistent with a previous, independent calibration for a similar sample of
spirals residing in clusters. Our analysis provides an accurate estimate of
the quantity $\beta_I\equiv\Omega_m^{0.6}/b_{I}$, where $b_{I}$ is the
linear biasing parameter for IRAS galaxies. Using the forward TF relation
and smoothing the predicted velocity field with a Gaussian filter of radius
300 \kmsn, we obtain $\beta_I=0.42 \pm 0.04$ (1-$\sigma$ uncertainty). This
value, as well as other parameters in the fit, are robust to varying the
smoothing radius to 500 \kms and splitting the sample into spherical shells
in redshift space. The one exception is the small-scale velocity
dispersion, $\sigma_v$, which varies from $\sim 200$
\kms within cz$_{LG}= 4000$ \kms to $\sim 500$ \kms at larger distance. For
$\beta_I \simeq 0.42$, the residuals between the TF data and the
PSC$z$ gravity field are uncorrelated, indicating that the model
provides a good fit to the data. More generally, a $\chi^2$ statistic
indicates that the PSC$z$ model velocity field provides an acceptable
($3\sigma$) fit to the data for $0.3<\beta_I<0.5$.

\end{abstract}




\newpage
\section{Introduction}
\label{sec:intro}

Measurements 
of peculiar motions 
provide a fundamental tool 
to probe the mass distribution in the local universe.  In the
linear regime of gravitational instability, a simple relation between
peculiar velocity, $\bv$, and mass density contrast, $\delta_m$, can be
easily obtained from 
mass conservation, either in differential
\begin{equation} 
\nabla \cdot \bv = -\Omega_m^{0.6} \delta_m,
\label{divv}
\end{equation}
or integral form,
\begin{equation}
\bv(\br)={{\Omega_m^{0.6}}\over{4\pi}}\int d^3\br \prime
{{\delta_m(\br \prime)(\br \prime-\br)}\over{|\br \prime-\br|^3}},
\label{intdelta}
\end{equation} 
where $\Omega_m$ is the cosmological mass density parameter.  Together with the
commonly used simplifying 
assumption of linear biasing, $\delta_g=b_g\delta_m$, where
$\delta_g$ is the galaxy density contrast and $b_g$ the galaxy biasing
parameter, these two equations provide a relation between
observable quantities: the peculiar velocity of luminous objects,
$\bv$, and the density contrast, $\delta_g$, which can be obtained from large
all-sky redshift surveys.
The IRAS Point Source Catalogue (Beichman \etal 1998)
is particularly suited for this purpose 
because of the good sky coverage and homogeneity of the 
survey. In this paper, we present a
comparison of peculiar velocity data with a redshift survey based on
the IRAS catalogue. Throughout the paper, we will use the subscript
$_I$ to indicate that an IRAS density field has been used.

Comparing $\delta_g$ with $\bv$ using any of the two equations above
allows us to estimate the quantity
$\beta_g={{\Omega_m^{0.6}} / {b_g}}$
and to test the validity of the gravitational instability hypothesis.
Although the two equations~(\ref{divv}) and~(\ref{intdelta}) 
 are mathematically equivalent, they lead to
two different strategies for measuring $\beta_g$.  
Equation  \ref{divv} is
used to perform the so-called density-density (d-d) comparisons which
typically consist of the following steps: 
a 3-D velocity field reconstruction from observed radial velocities; 
differentiation of $\bv(\br)$ and use of eqn.~(\ref{divv}) to compute
$\delta_g$;  
comparison to the observed galaxy density fields.  The first step is
the least trivial and requires some additional theoretical 
assumptions. 
Bertschinger \& Dekel (1989) successfully implemented a d-d technique
by assuming that $\bv(\br)$ is irrotational,
in the POTENT reconstruction method.
The many applications of POTENT to various datasets 
have consistently led to large values
of $\beta_I$ consistent with unity
(see Sigad  \etal 1998 
and references therein).
Equation~(\ref{intdelta}) is at the core of the so-called velocity velocity
(v-v) comparisons.  In this approach, one computes the mass density
field obtained from the galaxy distribution in the redshift survey,
uses equation~(\ref{intdelta}) to predict a peculiar velocity field and then
compares it with the observed galaxy velocities.  The v-v methods have
been applied to most of the catalogues presently available and have given
values of $\beta_I$ which are typically in the range $0.4 - 0.6$ (see
Willick 2000, for an updated summary of the various results).

The  v-v methods are commonly regarded as more reliable and robust
than the d-d
ones because they require less manipulation of the data.
The values 
of $\beta_I$ obtained by these analyses are significantly
smaller than unity, irrespective of the velocity tracers, model
gravity field and comparison technique used.  However, some of the v-v
analyses showed evidence for a poor match between models and data, which
would render  
the estimate of $\beta_I$ meaningless.  Davis, Nusser \&
Willick (1996) used their ITF technique (Nusser \& Davis 1994) to
compare the gravity field derived from the IRAS 1.2 Jy survey (Fisher
\etal 1995) with the peculiar velocities obtained from the Mark III
catalogue (Willick \etal 1995, 1996, 1997a).  The coherent dipole
residuals that they found were taken as evidence for significant
discrepancies between modeled and observed velocity fields.  Willick
\etal (1997b, V1 hereafter) and Willick \& Strauss (1998, V2
hereafter) also considered the IRAS 1.2 Jy velocity predictions and
the Mark III dataset but compared them using the VELMOD
method.  They were able to obtain a good fit to the data only by
introducing a physically motivated, external 
quadrupole 
contribution to the model velocity field.
Da Costa \etal 1998 (D98 hereafter) found good agreement
between the peculiar velocities of galaxies in the SFI catalogue 
(Haynes \etal 1999a, 1999b) 
and those derived from the IRAS 1.2Jy gravity
field, by performing an ITF comparison.  The same ITF method has been recently
applied to compare two different datasets: the IRAS PSC$z$ redshift
survey (Saunders \etal 2000) and the peculiar velocities in the
recently completed ENEAR catalogue (da Costa \etal 2000). Also in this
case the agreement between model and data was satisfactory
(Nusser \etal 2000).
 
In this paper, we use the VELMOD technique to compare the PSC$z$
velocity prediction to the SFI dataset. As for any v-v comparison our
main aim is to  constrain  $\beta_I$ 
and to investigate the adequacy of the
PSC$z$ model velocity field. However, rather than simply adding one more
measurement of $\beta_I$ to those already in the literature from other v-v
comparisons, we hope to address some more
specific questions which should help simplify the rather complicated 
picture that has emerged from the results of the various v-v
comparisons.  Our goal is to check whether the PSC$z$--IRAS gravity field still
provides a good fit to SFI data when the comparison is performed with
the VELMOD method rather than the ITF. 
We also want to exploit fully the denser and deeper PSC$z$ 
catalogue and see whether we can improve
the agreement between 
the gravity field and the measured velocities, thus
reducing the uncertainties in the estimate of $\beta_I$.

In Section~\ref{sec:method} we review the basics of the VELMOD
technique and describe its 
current implementation.  The SFI sample is
presented in Section~\ref{sec:sfi} 
and the  IRAS PSC$z$ catalogue and its
gravity field are described in Section~\ref{sec:pscz}.  VELMOD is tested 
in Section~\ref{sec:mockt} and the results of its application to the 
SFI catalogue  are 
presented in Section~\ref{sec:res}.  An analysis of
errors based on the magnitude and velocity residuals is performed in
Section~\ref{sec:errors}.  Finally, in Sections~\ref{sec:disc}
and \ref{sec:conc} we discuss the results and conclude.

\section{The VELMOD Method} 
\label{sec:method}

VELMOD is a maximum likelihood method introduced by V1 and described
in detail 
in 
V1 and V2.  Here we simply outline the main points
of the method, focusing on its implementation in the case of a forward
TF relation.  In Strauss \& Willick's (1995)
terminology, VELMOD
uses a Method II approach, i.e. takes the TF observables (apparent
magnitude and velocity width) and the redshift of an object and
quantifies the probability of observing the former given the 
latter, for a particular model of the velocity field and a TF
relation.  This probability is then maximized with respect to the
free parameters of the velocity model and the TF relation.  Unlike
previous Method II implementations (e.g. Hudson 1994), VELMOD replaces
the redshift-distance relation with a joint probability distribution
of distance and redshift.  This probabilistic approach allows a
statistical treatment of all those effects (small-scale velocity
noise, inaccuracy of the velocity model and existence of triple-valued
regions) that spoil the uniqueness of the redshift-distance mapping.

VELMOD does not require smoothing of the TF data which, along with the
allowance for triple-valued regions and small-scale velocity noise,
allows one to probe the velocity field in high-density regions, 
thus exploiting the denser sampling of the new PSC$z$ galaxy catalogue.
Another convenient  feature of VELMOD is that it does not require an {\it
a priori} calibration of the TF relation,
which is a common issue of concern 
in peculiar velocity studies.
Instead, a fit of the parameters of the TF relation
is performed simultaneously with the parameters of the velocity field.

\subsection{Implementation of VELMOD} 
\label{sec:methodimp}

For each galaxy, the angular position $(l,b)$, redshift
($cz$), apparent magnitude ($m$), and velocity width parameter ($\eta
\equiv \log_{10}(W) - 2.5$, where $W$ is twice the rotation
velocity of the galaxy) are taken from the SFI catalogue.
Hereafter, we will always use the redshift measured in the Local
Group (hereafter LG) frame
unless otherwise specified. In addition, a
model for the density field is needed to account for inhomogeneous
Malmquist bias.  Such a model is obtained from the distribution of IRAS
PSC$z$ galaxies, together with a self-consistent model for the peculiar
velocity field (as described in Section~\ref{sec:pscz}).  The velocity
model is completely specified by the value of the $\beta_I$ parameter and a
one-dimensional velocity dispersion, $\sigma_v$, which quantifies both
the inaccuracy of the velocity model and the true velocity noise
arising from small-scale nonlinear motions. Strauss, Ostriker \& Cen
(1998)  and V2
have shown that  the velocity dispersion on small scales is an
increasing function of the local density. The SFI galaxy sample
considered in this work consists of field spirals which avoid
high-density environments and thus we ignore such dependency and
assume a constant $\sigma_v$, independent of the environment.

The probability of observing a magnitude $m$ for a galaxy with a given 
$\eta$ and distance $r$ is modeled by a linear TF relation, 
\begin{equation}
m=M(\eta)+5 \log(r) = A_{TF} - b_{TF}\eta +5 \log(r) + \sigma_r,
\label{ftf}
\end{equation}
where $\sigma_r$ is a Gaussian random distribution with zero mean and 
dispersion $\sigma_{TF}$.  This TF relation is completely
specified by its zero point ($A_{TF}$), slope ($b_{TF}$) and scatter
($\sigma_{TF}$).  In this work we do not model a possible dependency
of the TF scatter on the luminosity or velocity width. In section
\ref{sec:vrot}, we will find that our results are insensitive 
to this approximation.

The final ingredient needed to  
compute the probability of observing the measurements of the SFI
catalogue are the selection function of the observational quantities
and their correlations.
Systematic errors can affect peculiar velocities computed with the forward
TF relation if selection effects are not propery accounted for.  
For that purpose, we have used the correlation and selection models of
Freudling \etal (1995).

With the above assumptions, one can compute the probability that the
$i$-th object of the sample with recession velocity $cz_{LG}$ and
velocity width 
parameter $\eta$ will have an apparent magnitude $m$: $P_i(m|\eta,
cz_{LG})$.  To evaluate this conditional probability one needs to
integrate the joint probability distribution $P_i(m,\eta, cz_{LG})$
over $m$. Although analytic approximations for this integral have been
introduced by V2, which are valid away from triple-valued regions and
at distances much larger than $\sigma_v$, in this work we 
perform an explicit numerical integration for all galaxies.

From the overall probability, which we obtain by multiplying 
the single-object probabilities $P=\prod_i P_i(m|\eta, cz_{LG})$, we
compute the likelihood ${\cal L} \equiv -2 \ln P$ which is then
minimized at each of 19 values of $\beta_I=0.1, 0.15 ..... 0.95, 1.0$
by continuously varying the remaining  free parameters (i.e. the
three TF parameters and $\sigma_v$).  The function
${\cal L}_{min}(\beta_I)$ obtained by this procedure is then fitted
with a cubic function and the 
maximum-likelihood value of $\beta_I$,
$\beta_{min}$, is found at the minimum of the curve.  Extensive tests
with mock catalogues performed by V1 have shown that $\beta_{min}$ is
an unbiased estimator of the true $\beta_I$ parameter.

V1 and V2 found systematic residuals between the IRAS 1.2Jy predicted
and the Mark III observed velocities which they modeled as a velocity
quadrupole with a distance-dependent amplitude.  This quadrupole,
which they treated as a free parameter in their VELMOD analysis, is
likely to arise from the missing contributions to the model velocity
field from the mass distribution in the regions beyond the limits of
the IRAS 1.2Jy survey and from shot noise.  The PSC$z$ survey
should be deep and dense enough to reduce such
discrepancies allowing us to drop this extra parameter.  Therefore, we
choose not to allow for any external contribution to the PSC$z$
velocity field in our VELMOD analysis. In Section~\ref{sec:errors} we
will show that the PSC$z$ velocity model within 6000 \kms
does indeed constitute an
acceptable fit to the SFI peculiar velocity field, with no need to
introduce external contributions.  Finally, unlike V1 and V2, we do not
model the uncertainties in the LG velocity as a free parameter.
Instead, we shall test explicitly the robustness of our results by
running a few VELMOD experiments in which the LG velocity is allowed
to vary within its 1-$\sigma$ error range.

\section{The SFI Sample} 
\label{sec:sfi}

The SFI sample of galaxies (Giovanelli \etal 1997a; Haynes \etal
1999a,b) is a homogeneous all-sky sample of galaxies for which I-band
Tully-Fisher parameters are available. The sample uses new
observations for declinations $\delta > -40^{\circ}$ and re-reduced data from
Mathewson, Ford and Buchorn (1992) in the southern polar cap. The sample is, by
design, angular diameter limited, with different limiting diameters, 
$D_{lim}$, for different redshift shells.  The limiting diameters,
expressed in unit of 0.1 arcmin, are $D_{lim} =25, 16$ and 13 in the
redshift ranges $cz_{LG}<3000 \kms$, $3000<cz_{LG}<5000 \kms$ and
$5000<cz_{LG}<7500 \kms$, respectively.  However, for a number of
different reasons, some bright galaxies smaller than the stated
diameter limit were included in the sample. 
These additional galaxies, which amount to $\sim 15\%$ of the total, 
do not constitute a strictly magnitude-limited sample. All of them are
brighter than an apparent Zwicky magnitude of
$m_z=14.5$. In a detailed investigation of
these extra galaxies, we have not found any dependence of the
selection on distance or apparent diameter. 
This is different from the
distance-dependent selection function that Willick \etal (1996) 
have used to describe the MAT sample.
For the present analysis, we have therefore approximated the SFI selection
sample as the combination of a strictly diameter-limited sample with
the above criteria, and a magnitude-limited sample with limiting
Zwicky magnitude of 14.5.  
For such a case, Willick (1994) has derived the
VELMOD formalism.  In his terminology, this 
is the {\it Two-Catalogue Selection} 
case  and the selection function, expressed in a form
suitable for the VELMOD analysis, is given by eq. (63) of Willick
(1994). 
We have found that the results presented in this paper 
do not change significantly when using $m_z=13.5$
for the putative magnitude limit of the extra bright galaxies

Freudling \etal (1995) have modeled the correlations between 
the quantities used to define the selection criteria,
$(m_z, D)$, and the TF observables, $(m_I,\eta)$, 
as 
\begin{equation}
m_z=1.6 + m_I +0.5 \eta \ \ \ \ \ {\rm with} \ {\rm dispersion} \
\sigma_{m_z}=0.59 \ 
\label{mz}
\end{equation}
and
\begin{equation}
D=3.82 + -0.205 m_I -0.102 \eta \ {\rm with} \ {\rm dispersion} 
\  \sigma_{D}=0.121,
\label{d}
\end{equation}
where $m_I$ indicates the I-band apparent magnitude.
These correlations can be easily translated into  the VELMOD formalism.

Below, we will consider the following sub-samples 
drawn from the SFI catalogue. Most
of the analysis is carried out with the 989 galaxies with
$cz_{LG}<6000 \kms$ and $\eta>-0.25$.  These  cuts
allow us to perform a homogeneous comparison with the ITF analysis of
D98. In addition, we will consider separate subsamples
restricted to three independent redshift-space shells 2000 \kms thick
and with external radii of 2000, 4000 and 6000 \kmsn. 
These subsamples
mix different nominal selection criteria and serve as an additional
check that selection effects are well accounted for. The three
subsamples contain 158, 355 and 496 galaxies, respectively.

SFI galaxies avoid rich clusters and high density environments.  In
particular none of them belong to the Virgo cluster. Therefore, there
is no need to adopt any grouping procedure and the hypothesis of a
small-scale velocity dispersion, $\sigma_v$, independent of the
environment is well justified.

\section{Model Density and Velocity Fields from the PSC$z$ Survey} 
\label{sec:pscz}

The models for the density and velocity fields are obtained from the
distribution of IRAS galaxies in the recently completed PSC$z$
all-sky redshift survey (Saunders \etal 2000).  This catalogue, which
basically extends the old 1.2 Jy one (Fisher \etal 1995), contains
$\sim 15500$ IRAS PSC galaxies with a flux at 60 $\mu m $ larger than
0.6 Jy.  A complete description of the dataset, its selection
criteria, the procedures adopted to avoid stellar contamination and
galactic cirrus, are given in Saunders \etal (2000).  For our
purposes, 
the most interesting features of the catalogue are the large area
sampled ($\sim 84 \%$ of the sky), its depth (the
median redshift is 8500 \kms) and dense sampling (the mean galaxy
separation at 10000 \kms is $\langle l \rangle \sim 1000$
\kmsn,\footnote {Throughout this paper we measure distances in velocity
units (\kms).  This is equivalent to setting the Hubble constant equal to
unity.}\ compared with a value of $ \langle l \rangle \sim 1500$
\kms for the IRAS 1.2 Jy catalogue).

The flux-limited nature of the catalogue causes the number of objects to
decrease with distance. This decrease is quantified by a 
radial selection function. Various authors have used different
estimators to compute the selection function (Springel 1996, Canavezes
\etal 1998, Sutherland \etal 1999, Branchini \etal 1999, Monaco and
Efstathiou 1999, Saunders \etal 2000).  Branchini \etal (1999,
hereafter B99) found that different selection functions induce 
variations smaller than 5\% in the model density and velocity fields 
within  a distance of 20000 \kmsn. In this work we use the selection function
specified by eqn.~(1) of B99.

The density and velocity field in real space are obtained from
the redshift space distribution of  PSC$z$ galaxies
by implementing   {\it Method 1} of  B99, which uses the 
iterative technique of Yahil \etal (1991)
to minimize redshift-space distortions.
The procedure relies on gravitational instability theory,
assumes linear biasing, and is valid in the limit of small density
fluctuations
where linear theory applies.  At each step of the iteration the 
gravity field, \bg, is computed  from the distribution of the 11206 
PSC$z$ galaxies within 20000 \kmsn.
The gravity field is subsequently smoothed with a top hat filter 
of  radius  $\sim 500$ 
\kms which allows us to assume linear theory and thus to obtain the 
smoothed peculiar velocity of each galaxy  from the acceleration,
$\bv \propto \beta_I \bg$.
The new distances of the objects, $r$, are then assigned assuming a unique 
redshift--distance relation  $ r = cz - u $, where $u$ is the radial
component of the peculiar velocity vector.
The procedure is repeated until 
convergence is reached.

The final products are the real space positions of the galaxies and their
velocities
for a given value of $\beta_I$. We have ran 19 different reconstructions
with $\beta_I = 0.1, 0.15 .....1.0$.
The continuous density field is obtained by smoothing the galaxy distribution
on  129$^3$ points of a cubic grid inside a box of 19200 \kms 
a side with the LG at the centre, using a Gaussian filter of 300 \kms
(G3, hereafter). The associated smoothed velocity fields are computed
for the appropriate value of $\beta_I$ from eqn.~(\ref{divv}).

The procedure returns 19
models of the density field (one for each value of $\beta_I$), 
along with their associated peculiar velocity fields, both defined at 
the gridpoint positions. Velocity predictions are made in the LG frame
to minimize the uncertainties derived from the lack of information about
the mass distributions on scales larger than the size of the PSC$z$
sample.

\section{Testing the VELMOD Implementation} 
\label{sec:mockt}

In this work we rely on the error analysis performed by V1, based on an 
extensive application of VELMOD to realistic mock catalogues, which 
we do not repeat here. Instead, we present the results of two tests aimed at 
checking the reliability of our VELMOD implementation.

\subsection{Tests with Ideal PSC$z$ and SFI Mock Catalogues}
\label{sec:mockt1}

As a  first test, we applied VELMOD to a suite of 
SFI and PSC$z$ mock catalogues. Errors in the  
VELMOD analysis are dominated by inhomogeneous Malmquist bias,
with uncertainties in the model velocity fields playing an important
role  only in the innermost part of the sample. 
To account properly for  Malmquist bias, 
the mass distribution in the mock catalogues needs to
mimic the one in our local universe as close as possible.
Unlike V1, however, we do not extract mock catalogues from
constrained N-body simulations (i.e. similar to those
produced by Kolatt \etal 1996). 
Instead,  we construct 'ideal' 
PSC$z$ and SFI mock catalogues from the G3 smoothed PSC$z$ density 
and reconstructed linear velocity fields
(described in the previous section) for $\beta_I=1.0$.
Nonlinear effects were mimicked by adding a 
constant Gaussian noise of  150 \kms to 
each  of the  Cartesian components
of the model velocity vectors. 
Such a small value is meant 
to mimic the observed ``coldness'' of the 
cosmic velocity field  (e.g. Strauss, Ostriker and Cen 1998).

Twenty mock PSC$z$ catalogues where generated with Montecarlo techniques
using different 
random seeds. Selection criteria close to the observational ones
were applied to mimic the
PSC$z$ selection function and the presence of unobserved regions.
Mock PSC$z$ galaxies were generated assuming a probability
proportional  to the local density, i.e. assuming that they trace the mass
distribution ($b_I=1$).
The same reconstruction method used for the real data was then applied to 
each of these mocks to obtain 20 mock PSC$z$ velocity models 
for each of the 20 values of $\beta_I=0.5, 0.55.....1.5$.

Similarly, we generate 20 mock SFI catalogues by Montecarlo resampling 
the original PSC$z$ G3--smoothed density and velocity fields with the
added thermal noise. First,  catalogues with large numbers  of galaxies were
created.  Next, absolute magnitudes  were assigned according to the luminosity 
function. 
Subsequently,  diameters and velocity widths were assigned according to the 
correlations described in Section~\ref{sec:sfi}. TF parameters  close to the
ones found in our final analysis were used to assign velocity widths. Finally, 
the nominal selection criteria of the SFI sample were used to select galaxies 
for the final catalogues.  The additional galaxies mentioned in
Section~\ref{sec:sfi} were simulated by randomly including galaxies up
to the magnitude limit until the number of galaxies not satisfying the
nominal selection criteria matched the one of the SFI sample.  The
resulting mock samples nicely reproduce the observed redshift
distribution of the SFI sample. 
 

The results of applying VELMOD to the 20 SFI mock catalogues are 
summarized 
in table~\ref{tab:mock}.  The true values of the free parameters used in
the mock catalogues are listed in the first row while those obtained
from the VELMOD analysis are shown in the second row. For each
parameter we report the average value from the 20 mocks, the error in
the mean and, in parenthesis, the typical error in a single
realization.  These results indicate that VELMOD returns an unbiased
estimate of the free parameters.

Given the ideal nature of the velocity field in these 
mock catalogues the errors displayed in 
table~\ref{tab:mock} are likely to underestimate
the real ones.
In what follows, however, we will be mainly interested in the errors 
on  $\beta_I$  which, as V1 demonstrated, 
can be obtained from the values at which  
${\cal L}_{min}(\beta_I)$ differs by one unit from its minimum value
at  $\beta_{min}$.

\subsection{Tests with the Mark III Catalogue}
\label{sec:mockt2}

In the second test, we repeated part of the VELMOD analysis 
performed by
V1 and V2, but using the PSC$z$ model velocity field instead of the IRAS 1.2
Jy one.
We applied VELMOD to the Aaronson \etal (1982, hereafter A82) and
Mathewson \etal  
(1992, MAT) 
subsamples selected  according to the V1 prescriptions.
The selection functions  for A82 and MAT and their coefficients were taken
from 
Willick \etal (1996). Both samples are spatially limited to a redshift
$cz_{LG}=3000 \kms$. A third subsample
we use,  MAT2, also obtained from the Mathewson
\etal (1992) catalogue, coincides with the one 
considered by V2 and extends out to $cz_{LG}=7500 \kms$.

The results are summarized in  table~\ref{tab:mk3} where the values 
of the free parameters obtained from our VELMOD analysis are compared
to those obtained by V1 and V2 (in parenthesis). Only errors in  $\beta_I$
are quoted. Our results generally agree with those of  V1 and V2.
The analysis of the MAT and MAT2 samples 
returns values of $\beta_I$  smaller 
than those obtained by V1 and V2. The 
difference is within the 1$\sigma$ error bar
and probably reflects the difference between the
IRAS 1.2 Jy and PSC$z$ model velocity fields.

The largest discrepancy is  the $\sigma_v$ parameter for
which we consistently obtain values 
larger by $30-70$\%  than those estimated by V1 and V2.
Such discrepancy is statistically significant since
the expected errors on $\sigma_v$ estimated from the mock catalogues and, 
more qualitatively, from the
likelihood curve, are of the order of 40 \kmsn.
To interpret this discrepancy one has to keep in mind that
the value of $\sigma_v$ is approximately given by the sum in
quadrature of random velocity noise ($\sigma_T$) 
and the uncertainties in the model velocity field ($\sigma_M$).

The only possible explanation for the discrepancy in $\sigma_v$ is that
$\sigma_T$, which is an intrinsic property of the velocity catalog, 
is the same in both analyses, but the PSC$z$ velocity model
has a larger $\sigma_M$.  Indeed, since B99 estimate $\sigma_M \sim 130$
\kms for the PSC$z$ model, while V1 find $\sigma_M \sim 84$ \kms for
the IRAS 1.2Jy model, one may think that the intrinsic velocity noise
would then be similar in the PSC$z$ ($\sigma_T \sim 120$
\kmsn) and IRAS 1.2Jy ($\sigma_T\sim 100$ \kmsn).
However, it is difficult to understand why errors in the PSC$z$ model
should be larger than in the IRAS 1.2Jy one. The denser sampling and
larger depth of the PSC$z$ catalogue should decrease the uncertainties
in the model velocity field, not increase them. In fact, the errors
quoted by V1 and B99 were estimated using different mock
catalogues. V1 used mocks derived from the reconstruction technique of
Kolatt \etal (1996) which are based on a PM N-body code and are
intrinsically ``colder `` than those used by B99 which were taken from
the higher resolution AP$^3$M N-body simulations of Cole \etal (1998).
Since the precision of the velocity reconstruction decreases with
increasing small-scale random velocity noise, 
it is reasonable to suspect that
the difference in  the values of $\sigma_M$ estimated by V1 and 
B99 simply reflects the differences in the mock catalogues rather 
than differences in the errors of the two model velocity fields,
and that an error analysis based on the same set of mock catalogs
would show that the value of $\sigma_M$ for the PSC$z$ model is comparable, 
if not smaller, than for the IRAS 1.2Jy model velocity field.

Nevertheless, the $\sigma_v$ discrepancy must reflect some
difference in the IRAS PSC$z$ and 1.2Jy velocity prediction.
The only remaining possibility is that such a difference 
is systematic rather than purely random.
Indeed, the two models are different because they
are based on two different redshift catalogues and because they are
derived using two different reconstruction methods.  The PSC$z$
catalogue allows a denser sampling and therefore traces small-scale
density fluctuations which are washed out by shot noise in the
IRAS 1.2Jy catalogue.  The IRAS 1.2 Jy model used by V1 and V2 was
obtained using the technique of Sigad \etal (1998) which differs from the
one of B99 in the treatment of triple-valued regions, the use of Wiener
filtering, and, most importantly, in the allowance for mildly
nonlinear motions.  As a result, the PSC$z$ velocity model is
intrinsically more linear than the IRAS 1.2Jy one which also lacks 
power on small scales.  The linearity of the PSC$z$ model forces any
discrepancy between true and model velocities to contribute to
$\sigma_v$.  In the IRAS 1.2Jy case, such discrepancies are, on average,
smaller because of the lack of small-scale power and because they are 
partially absorbed by the mildly nonlinear motions.  The net result is
that VELMOD analyses based on the IRAS 1.2 Jy velocity model return
values of $\sigma_v$ systematically smaller than those based on the
PSC$z$ velocity model and a value of $\beta$ slightly larger.

\section{Results} 
\label{sec:res}
In this section we present the results of applying 
the VELMOD analysis
to the SFI sample within $6000$ \kms and we check the robustness of our 
results using smaller subsamples, smoothing scale and errors in the LG 
velocity.

\subsection{Application to the Full  SFI Catalogue}
\label{sec:full}

\begin{figure}
\vspace{11.0truecm}
{\includegraphics{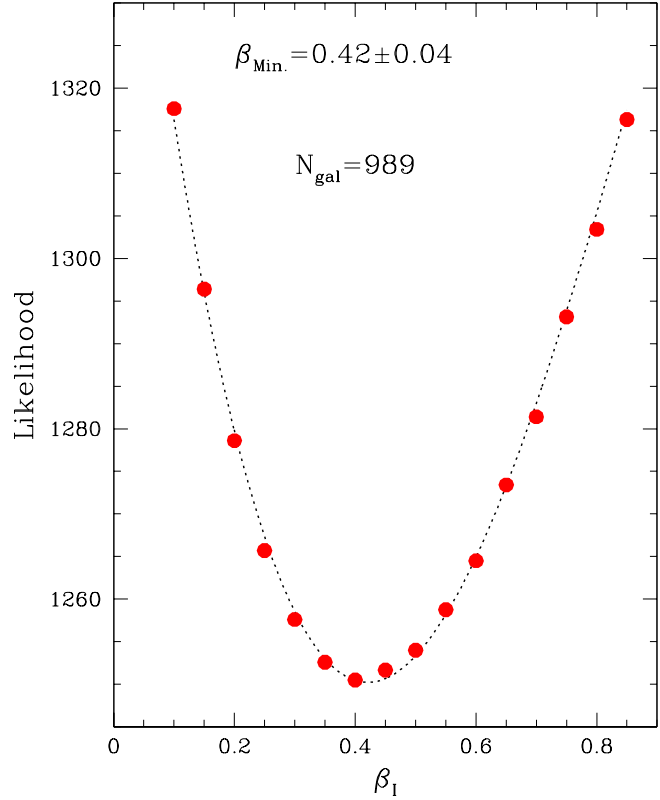}}
\caption{\capt The VELMOD likelihood , ${\cal L}_{min}(\beta_I)$,
(filled symbols) for the full SFI sample.
The dotted  curve represents a cubic fit to the points.}
\label{fig:Vfull}
\end{figure}

We ran VELMOD on the full SFI sample of 989 galaxies 
with redshift $cz_{LG}<6000$ \kms and $\eta>-0.25$.
The PSC$z$  density and velocity  
fields used in this calculation were smoothed
with a Gaussian filter of effective radius 300 \kmsn.
The likelihood was minimized at each of 19 values of $\beta_I=0.1,
0.15,....1.0$
by varying the free parameters $A_{TF}$, $b_{TF}$, $\sigma_{TF}$ and
$\sigma_{v}$.
The resulting values of ${\cal L}_{min}(\beta_I)$ are represented by
filled dots  in  fig~\ref{fig:Vfull}.
In order to find the actual minimum, we fit a cubic function  
to ${\cal L}_{min}(\beta_I)$:
\begin{equation} 
{\cal L}_{min}={\cal
L}_0+a_1(\beta_I-\beta_{min})^2+a_2(\beta_I-\beta_{min})^3
\label{cubic}
\end{equation}
The value of $\beta_{min}$ and its errors are indicated in the plot
and are also listed in table~\ref{tab:res} together with the values of the
other parameters found at the minimum of the likelihood curve.

The values for the TF parameters listed in table~\ref{tab:res} 
should be compared with the values 
found by Giovanelli \etal (1997b, G97) 
for an associated sample of cluster galaxies, termed SCI, 
selected in a similar way to the SFI sample,
which are commonly used as the assumed TF relation for the SFI sample. 
They find zero point, $A_{TF}$, of 
$-6.00\pm0.02$\footnote {In this work absolute magnitudes refer 
to distances in units of \kmsn, and not of 10 pc, as often adopted in the 
literature}\ 
and slope, $b_{TF}$, of $7.5\pm 0.2$ with a scatter of about 0.36 magnitudes. 
While the range of values for the slope of this determination 
from cluster galaxies overlaps with the current results,
we obtain a smaller value for the zero point and a larger scatter. 
This is in agreement with other analyses 
of the SFI dataset (D98, Freudling 
\etal 1999). A similar difference in the TF 
scatter of cluster and field samples has been found in other datasets
(e.g. Bothun \& Mould 1987; Freudling, Martel \& Haynes 1991).

The recovered value of $\sigma_{v}=250$
\kms is twice as large as the value obtained by V1 and V2.  As
we have already discussed, 
part of this discrepancy may simply reflect
differences between our model velocity field and the one used by V1
and V2.  It may also indicate that the
PSC$z$ velocity model does not provide a satisfactory fit to the SFI
dataset.  We will return to this important point in
Section~\ref{sec:errors}.

\begin{figure}
\vspace{11.0truecm}
{\includegraphics{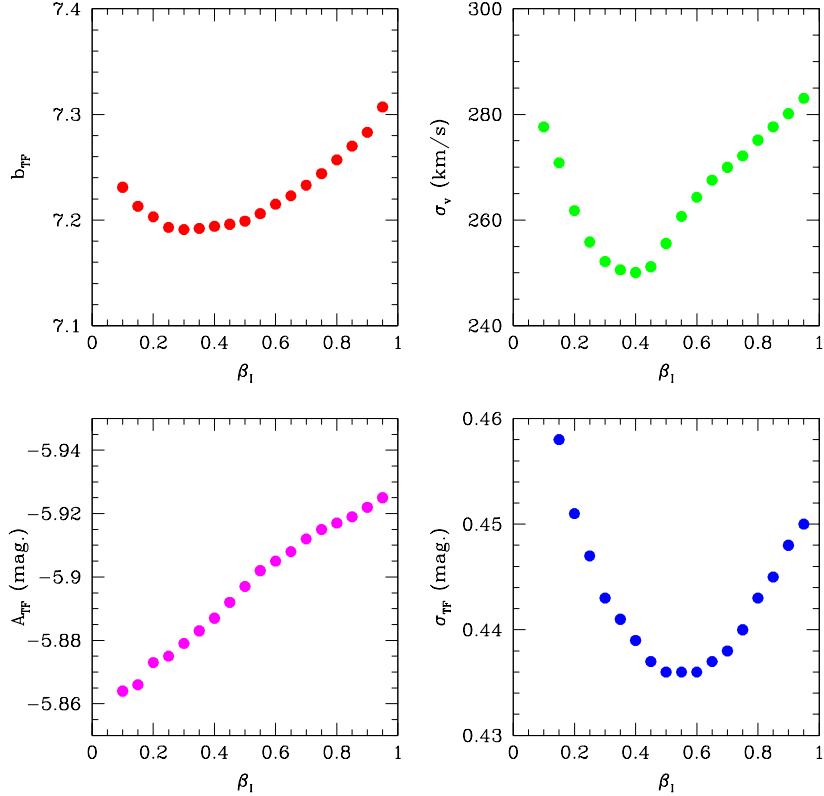}}
\caption{\capt Best fit parameters obtained from running VELMOD on the full sample
as a function of $\beta_I$: TF slope (top left), TF zero point (bottom left)
velocity dispersion (top right) and TF scatter (bottom right).
}
\label{fig:par}
\end{figure}

Figure \ref{fig:par} shows the variation of the four parameters with
$\beta_I$, where the other parameters are allowed to vary.
The velocity dispersion reaches a  minimum when
$\beta_I \sim \beta_{min}$, a behavior consistent with the analysis of the mock
catalogues performed by V1 (but not with their analysis of the real A82+MAT
sample).  The TF parameters are 
very insensitive to $\beta_I$. The
largest variation ($\sim$ 5\%) occurs for the TF scatter which has a
minimum around $\beta_I =0.55$, significantly larger than
$\beta_{min}$.  
This is not alarming since, 
as stressed by V1, 
minimizing the TF scatter does not correspond to maximizing the likelihood.
This is probably 
due to the covariance between $\sigma_{v}$ and
$\sigma_{TF}$: the increase of $\sigma_{v}$ after its minimum is
compensated by a further decrease of $\sigma_{TF}$.

\subsection{The Effect of Smoothing}
\label{sec:smooth}

In their analysis, V1 noticed that smoothing the model velocity field
with a G5 filter had little effect on the VELMOD results.  Along with
their small value of $\sigma_{v}$, this was taken as an indication that
density fluctuations on scales between 300 and 500 \kms contribute
little to both the Mark III and the IRAS 1.2Jy velocity fields.
However, the larger value of $\sigma_{v}$ we have found in our VELMOD
analysis of the A82, MAT and SFI samples seems to indicate that
contributions from small scales are non-negligible. 

To investigate this issue, we have repeated the same exercise and
performed a VELMOD analysis using a G5 smoothed PSC$z$ model velocity
field instead of the G3 one.  We find that the results are insensitive
to the smoothing scale, with $\beta_{min}$ increasing only by $\sim
5$\%, consistent at the 1$\sigma$ level with the value found in the G3
case.  The same is true for the other free parameters with the
exception of $\sigma_{v}$ which increases by $\sim 15$\%.  Moreover,
the value of ${\cal L}_{min}$ at $\beta_{min}$ increases by 10 units
with respect to the G3 case, corresponding to a probability decrease
by a factor $e^5\simeq90$ (as ${\cal L}=-2\ln P$).  How do we
interpret these results ?  With a G5 filter, the model velocity field
does not receive contributions from scales smaller than $\sim500$
\kms and larger values of $\beta_I$ and $\sigma_{v}$ are needed to
match the amplitude of the observed velocities.  This means that the
linear PSC$z$ model velocity field does receive a non-negligible
contribution from small scales.  The significant increase in the
likelihood indicates that the G3 velocity field provides a better fit
to the SFI velocities, i.e. small-scale fluctuations do contribute to
the peculiar velocities of galaxies in the real world.

\subsection{Uncertainties in the Velocity of the Local Group}
\label{sec:lg}

Since model predictions are given in the LG frame (i.e.  the
LG velocity is subtracted from all other velocities), we need to
quantify the impact of uncertainties in the predicted and measured LG velocity
when performing the VELMOD analysis.  V1 and V2 tackled the problem 
by introducing a free parameter to model a
random component to the LG velocity vector.  Here we take a different
approach aimed at minimizing the number of free parameters in 
the analysis.
Instead of modeling the uncertainties in the LG
velocities, we quantify the impact that these uncertainties,
independently estimated, may have on our VELMOD analysis.  
Errors in the LG velocity derive from uncertainties in modeling the LG
velocity and in transforming redshifts from the heliocentric to
the LG frame. The former have been quantified by Schmoldt \etal
(1999), while for the latter we consider the recent work of
Courteau \& van den Bergh (1999). 
We assume that these error estimates 
are independent and so we compute the
total uncertainty by adding them in quadrature.  The resulting total
error in the LG velocity is $\sim$120$\beta_I$ \kms for each Cartesian
component.

To evaluate the impact of these errors we ran 10 VELMOD analyses in
which the LG velocity was perturbed with a Gaussian noise of the
same amplitude.  The effect is to increase the random errors
without introducing any systematic bias.  Averaging over the 10
experiments we obtain a value of $\beta_I=0.43$ with a 1-$\sigma$
error around the mean of $0.05$, slightly larger than the typical
error in a single realization.  Similar considerations apply to the
other parameters.  In all but one experiment the value of ${\cal
L}_{min}(\beta_{min})$ was larger than in the unperturbed case. In
the one case with smaller likelihood (by only two 
units) the perturbation
turned out to be very small, in agreement with the V1 and V2 results
in which the extra random components added to the LG velocity
vector turned out to be trivially small.
 
\subsection{Breakdown by Rotation Velocities}
\label{sec:vrot}

In their calibration of the TF relation for SCI galaxies, G97 found
that the TF scatter is a function of the velocity width: rapidly
rotating galaxies have a smaller TF scatter than slow rotators.  A
similar trend has also been detected by Federspiel, Sandage \& Tamman
(1994) and by Willick \etal (1997a) 
in some of the Mark III
subsamples. In the SFI catalogue itself this effect has been
included for the direct TF relation by Freudling \etal (1999), but not
in the inverse TF relation (D98). In the current analysis, we have
chosen to neglect this effect.  To estimate the impact 
of this approximation 
we divided the SFI catalogue into two subsamples according to the
rotational velocity of the galaxies.  In the first catalogue, we
included 
only objects with a velocity width parameter $\eta>-0.1$,
while in the second we included slow rotators with
$\eta<-0.1$.  The results of applying VELMOD on these two
samples are shown in table~\ref{tab:res}.  The TF dispersion is found
to be smaller for fast rotators,
as could be expected.
However, the values of the other free parameters, 
in particular $\beta_{min}$, 
are very similar in the two subsamples,  showing that our approximation of 
a constant $\sigma_{TF}$ has little impact on our $\beta_I$ estimates.

\subsection{Breakdown by Redshift}
\label{sec:reds}

\begin{figure}
\vspace{11.0truecm}
{\includegraphics{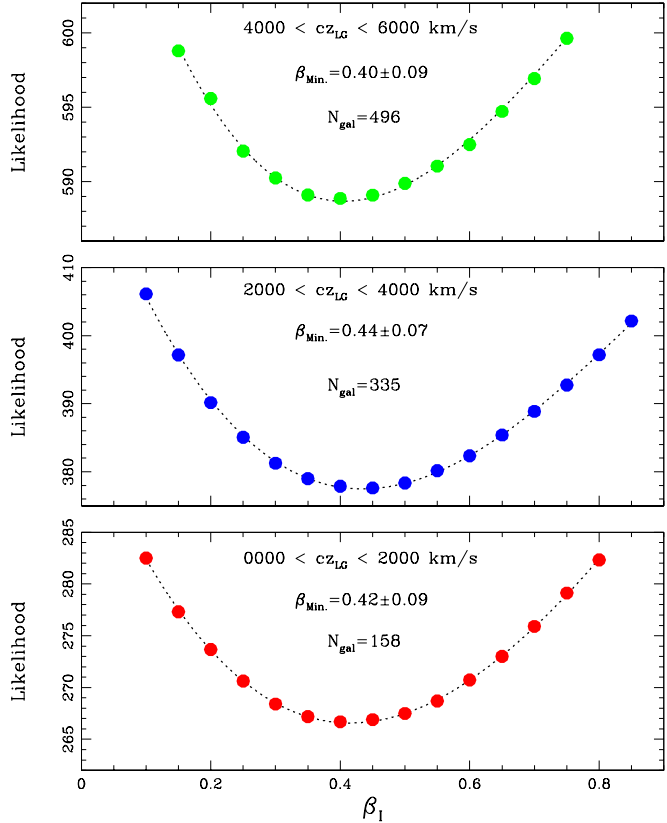}}
\caption{\capt Likelihood curves, ${\cal L}_{min}(\beta_I)$, for three
different 
redshift intervals. In each case the value of $\beta_{min}$ and the
number of 
galaxies in the samples are  indicated.
The dotted curves are cubic fits to the ${\cal L}_{min}(\beta_I)$ points.
}
\label{fig:zshell1}
\end{figure}

As a last robustness test we have divided our sample into three
independent redshift shells 2000 \kms thick and applied VELMOD to each
of them. Since the selection criteria are different in the three
shells, this test serves as a check of whether selection effects are
properly taken into account by our procedure. The resulting ${\cal
L}_{min}(\beta_I)$ curves are displayed in fig.~\ref{fig:zshell1}
and the free parameters are listed in table~\ref{tab:res}. The values
of $\beta_{min}$ in the three shells are 
consistent
with each other. The remaining parameters are also in good agreement,
with only two exceptions.  One is the TF slope which in the innermost
shell is $\sim 10$\% shallower than in the rest of the sample.  The
analysis of SFI mock catalogues in Section~\ref{sec:mockt1} reveals that
such a discrepancy is significant at the 1.5$\sigma$ level. However, as
we have already pointed out, the errors estimated using those catalogues
are smaller than the real ones. A VELMOD analysis of more realistic
SFI mocks would return larger errors and decrease the statistical
significance of the slope variation.

\begin{figure}
\vspace{11.0truecm}
{\includegraphics{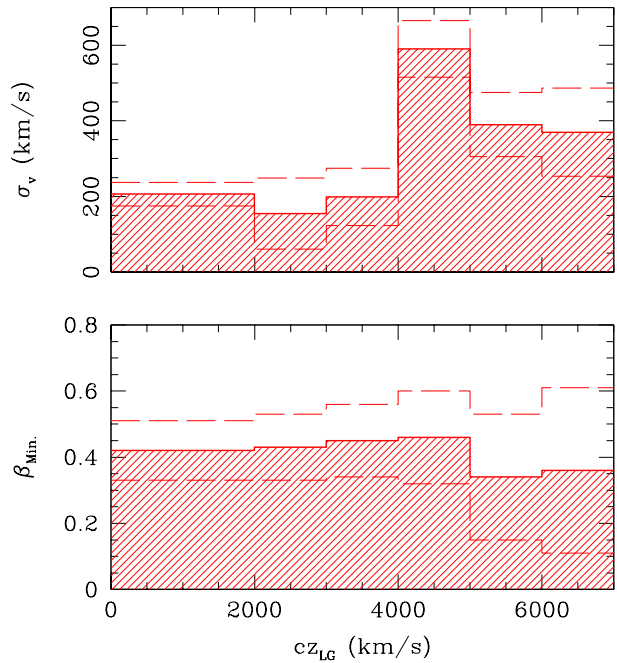}}
\caption{\capt Dependency of $\sigma_v$ (upper plot) and
$\beta_{min}$ (lower plot)
on redshift. The histograms show the results of running VELMOD on 6
different 
redshift shells. In both plots the dashed lines represent 1$\sigma$
errors around the mean. }
\label{fig:zshell2}
\end{figure}

The second, more serious, discrepancy is observed for $\sigma_v$ which
in the outermost shell increases by a factor of $\sim2.7$.  
Such a
variation is much larger than what we observe in the mock catalogues and
deserves some further investigation. We have therefore sliced the SFI
sample into thinner redshift shells, all of them 1000 \kms thick except
the innermost one which we extended to 2000 \kms so that a
comparable number of objects is contained in each shell.  We then 
repeated the VELMOD analysis in each of these shells.  The
results are visualized in the histograms of fig~\ref{fig:zshell2}.
The two dashed lines represent the 1$\sigma$
error about the mean. 
The upper plot shows the radial behaviour of $\sigma_v$.  Within 4000
\kmsn, $\sigma_v\simeq 200$ \kmsn, comparable to what we have obtained
from the analysis of the A82, MAT and MAT2 samples.  This value is
larger than those found by V1 and V2 and, as discussed in
Section~\ref{sec:mockt2}, this is probably because the 
the PSC$z$ model velocity field is more sensitive to power on small
scales.  Around 5000 \kms $\sigma_v$ increases to $\sim 600$ \kms and
then stabilizes at a value of $\sim 400$ \kmsn. 

In order to test whether any particular part of the sky is responsible
for this dependence of $\sigma_v$ on distance, we have cut the sample
into several complementary hemispheres (e.g. above and below the the
Galactic plane, above and below the Supergalactic plane and so on)
and, for each pair, we have ran VELMOD on the two sets of
redshift shells.  In all cases we have found that $\sigma_v$ increases
beyond 4000 \kmsn, suggesting that no particular cosmic structure is
responsible for the increase in $\sigma_v$.

The ability of VELMOD to constrain $\sigma_v$ is a rapidly
decreasing function of the redshift and we need to assess the
statistical significance of the jump at 4000 \kmsn, i.e. we
need to estimate the errors on $\sigma_v$.  We compute them from the
distribution of $\sigma_v$ found from the analysis of our 20 mock SFI
samples.  This test reveals that the increase of $\sigma_v$ is
significant at about the 3$\sigma$ level, leaving little doubt about
the reality of the change at around 4000 \kmsn.  
However, as we have already stressed, such tests
are only indicative since non-linearities are not properly modeled in
our mock samples. 

To further assess the reality of this peak we 
performed a second test.  We ran two VELMOD analyses using only
objects with redshifts in the range 4000--6000 \kmsn. In the two
calculations all the parameters were fixed to their maximum likelihood
values except $\sigma_v$ which was set equal to 200 \kms in one case
and to 535 \kms (i.e. its maximum likelihood value) in the other.
The smaller  
$\sigma_v$ results in an increase of 12  units in the
likelihood or a decrease in probability of $\sim e^6\simeq 400$,
indicating that the increase in $\sigma_v$ seen in the external shell
is indeed significant.  It is therefore possible that the dramatic
increase in $\sigma_v$ reflects a disagreement between the velocity
model and the SFI data on large scales. We will investigate this
possibility further in the next section.

\begin{figure}
\vspace{11.0truecm}
{\includegraphics{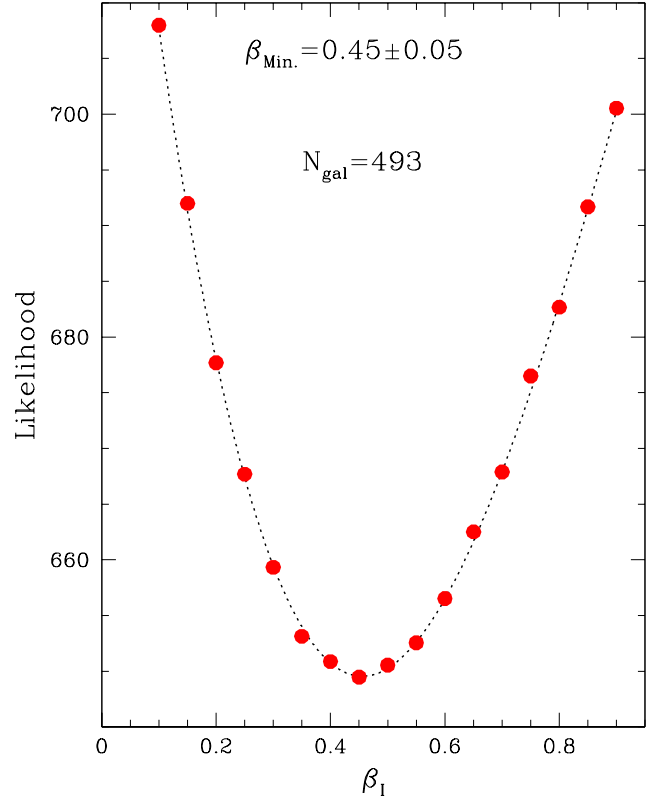}}
\caption{\capt The VELMOD likelihood, ${\cal L}_{min}(\beta_I)$, 
(filled symbols) for  SFI galaxies with $cz_{LG}<4000$ \kmsn.
The dotted  curve represents a cubic fit to the points.}
\label{fig:Vfull2}
\end{figure}

The lower plot of fig~\ref{fig:zshell2} shows how $\beta_{min}$
varies with redshift.  VELMOD returns a very robust estimate of
$\beta_{min}$ which does not change significantly even when $\sigma_v$
increases.  Despite this encouraging evidence, it is sensible to
adopt a more conservative approach and repeat the VELMOD analysis
within 4000 \kmsn, i.e. in the volume where there are no obvious
indications of a 
possible mismatch between model and data.  The results, listed in
table~\ref{tab:res}, are in good agreement with those of the full
sample. In particular, the minimum of the likelihood curve (shown in
fig.~\ref{fig:Vfull2}),
$\beta_{min}=0.45\pm0.05$, is fully consistent with the results from
the other subsamples.  The velocity dispersion is $\sim 200
\kms$, in good agreement with those obtained from the analysis of the
A82, MAT and MAT2 samples.

\section{Analysis of the Velocity and Magnitude Residuals}
\label{sec:errors}

VELMOD is a maximum likelihood technique in which the sources of
variance, $\sigma_v$ and $\sigma_{TF}$, are treated as free
parameters. For this reason, the VELMOD analysis can only tell us which
are the best values of $\beta_I$, $A_{TF}$, $b_{TF}$, $\sigma_{TF}$
and $\sigma_{v}$ for a given velocity field model, but it cannot
address the question of whether the velocity model is an acceptable fit
to the data.  In this respect, the 
increase of $\sigma_{v}$ at large radii found in the preceeding section 
is only suggestive of a mismatch between
observed and modeled velocities, and a proper error analysis is needed
to assess its statistical significance.  In this section we address
this problem by inspecting the magnitude and velocity 
residuals, following the formalism and notation of V1. 

\subsection{Maps of Velocity Residuals}
\label{sec:errormap}

\begin{figure}
\vspace{11.0truecm}
{\includegraphics{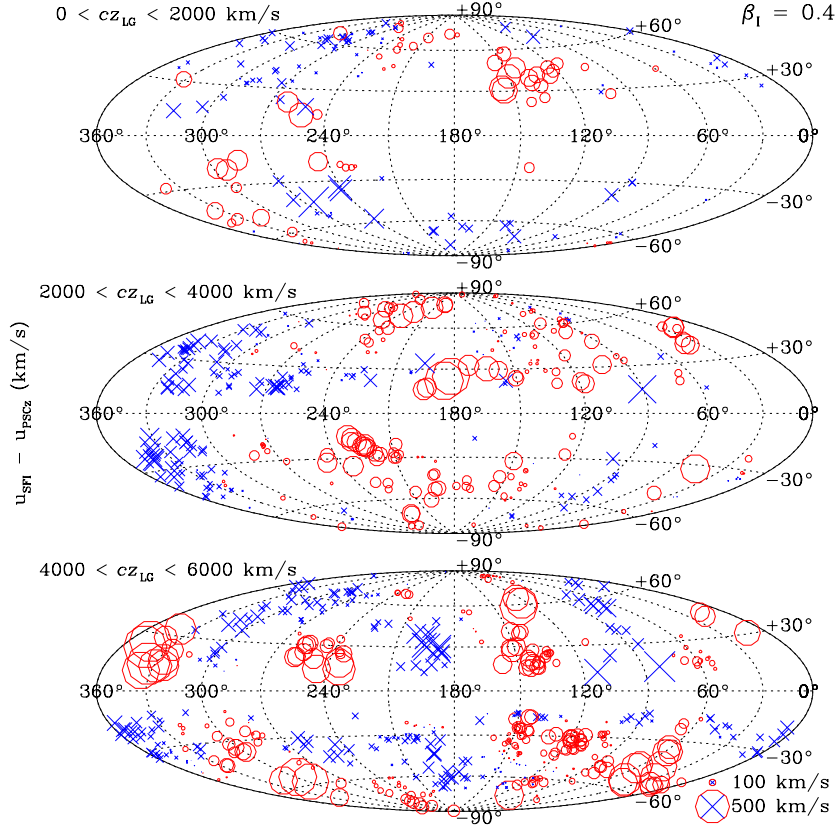}}
\caption{\capt The sky projection in Galactic coordinates, as seen in the
LG frame, of the smoothed VELMOD velocity residuals, $u_{SFI}-u_{\rm
PSC{\it z}}$, for $\beta_I=0.4$. Open circles indicate objects that
are inflowing relative to PSC$z$ velocity predictions; crosses denote
objects that are outflowing. The size of the symbols is proportional
to the amplitude of the velocity vector.  }
\label{fig:vmap04}
\end{figure}

\begin{figure}
\vspace{11.0truecm}
{\includegraphics{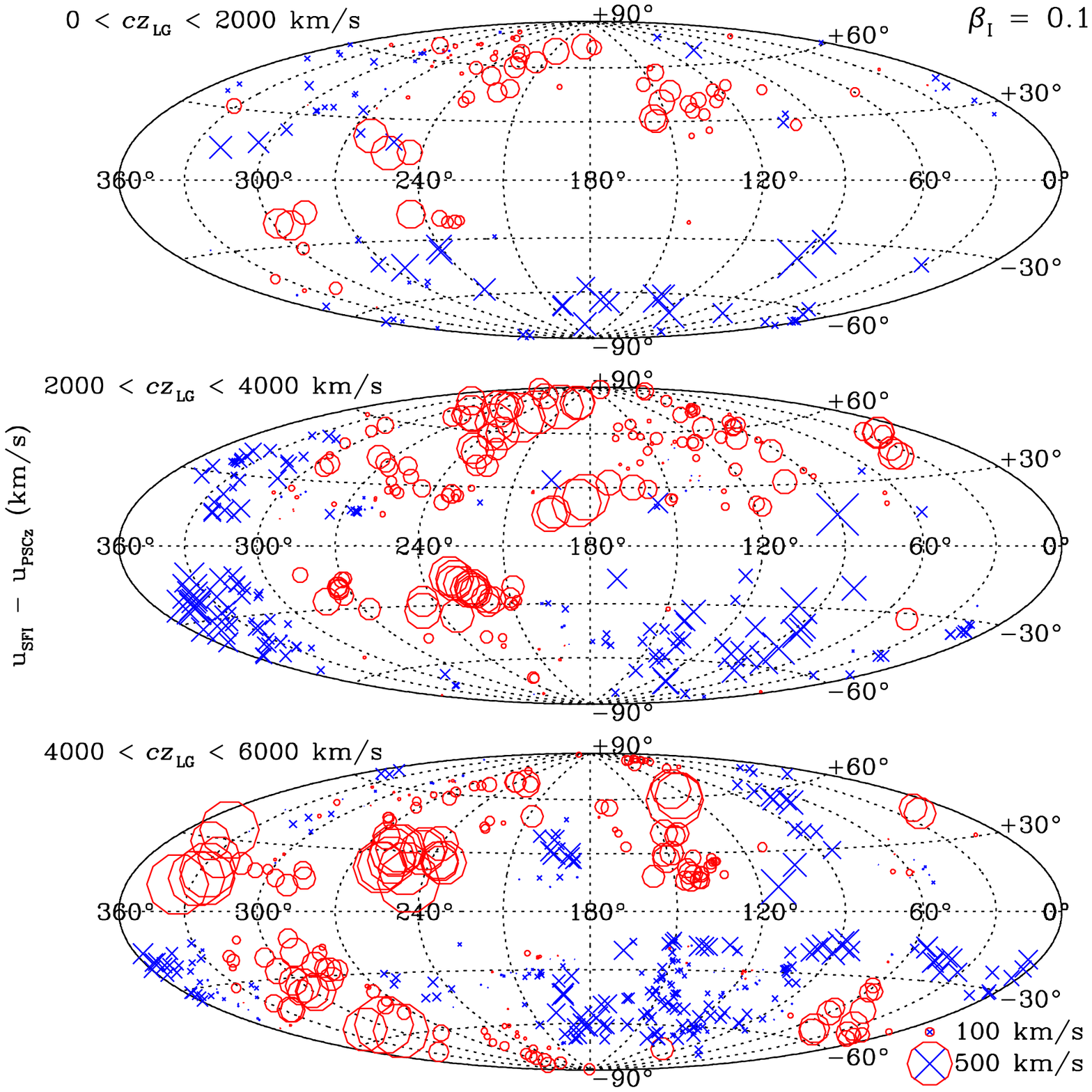}}
\caption{\capt The sky projection of the smoothed VELMOD velocity
residuals 
 $u_{SFI}-u_{\rm PSC{\it z}}$  for  $\beta_I=0.1$.}
\label{fig:vmap01}
\end{figure}

\begin{figure}
\vspace{11.0truecm}
{\includegraphics{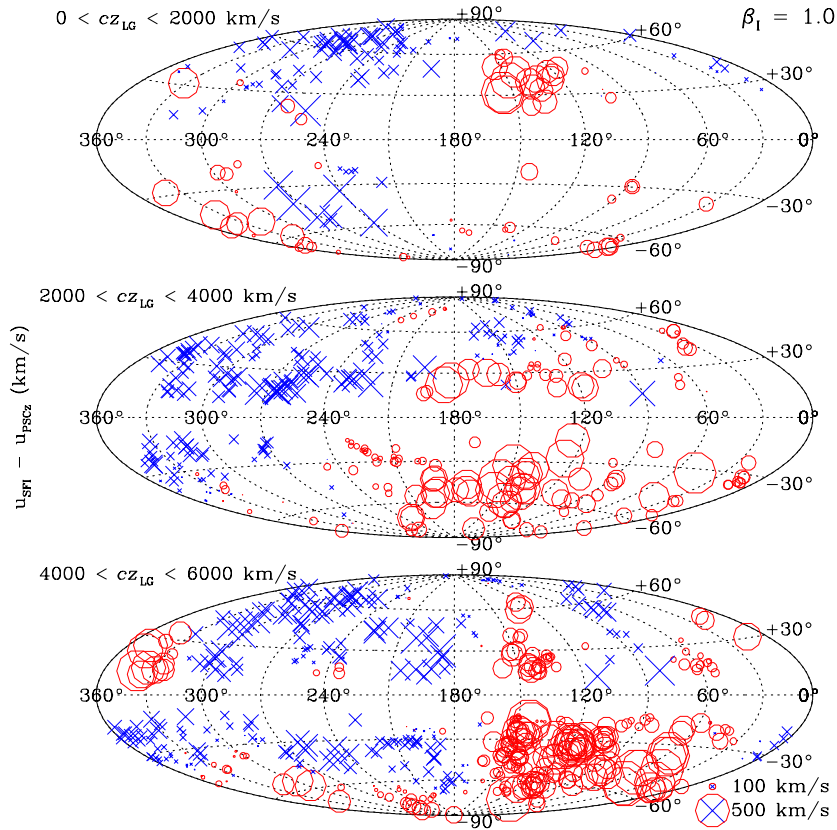}}
\caption{\capt
The sky projection of the smoothed VELMOD velocity residuals, 
$u_{SFI}-u_{\rm PSC{\it z}}$, for  $\beta_I=1.0$
}
\label{fig:vmap1}
\end{figure}

We define the normalized magnitude residuals for each object
of magnitude $m$:
\begin{equation} 
\delta_m={{m-E(m|\eta,cz)}\over{\Delta_m}},
\label{magres}
\end{equation}
where the expected apparent magnitude, $E(m|\eta,cz)$, and the
dispersion around it, $\Delta_m$, can be obtained by integrating over
the constrained probability function, $P(m|\eta, cz_{LG})$, which is
computed in the VELMOD analysis.  The magnitude residuals where
smoothed on a scale $S=d/5$, where $d$ is the PSC$z$ predicted
distance of the generic object, and then converted into smoothed
peculiar velocity residuals, $u_{SFI}-u_{\rm PSC{\it z}}$, according
to the V1 prescriptions.  The resulting maps of the velocity residuals
are shown in figs.~\ref{fig:vmap04},
\ref{fig:vmap01} and ~\ref{fig:vmap1}
for three different values of $\beta_I$ and in three different
redshift shells.  When looking at the maps one should bear in mind
that coherence up to scales of $\sim 35^{\circ}$ is induced by the
adopted smoothing and therefore systematic mismatches between 
the model and reality can only be
revealed by coherence on much larger angular scales.  The
visual inspection of the maps clearly shows that the velocity
residuals for $\beta_I=0.4$ are less coherent and have smaller
amplitudes than those of the models with $\beta_I=0.1$ and
$\beta_I=1.0$.  The residuals in the $\beta_I=0.1$ map show a pattern
similar to $\beta_I=0.4$ case, but have a larger amplitude and
coherence.  When $\beta_I=1.0$, the residuals grow even larger and the
map exhibits a clear dipolar structure.  The velocity residuals of the
model
with $\beta_I=0.4$ look qualitatively similar to the ones between the
SFI and IRAS 1.2Jy velocities for $\beta_I=0.6$, as seen in fig.~7 of
D98. However, a one-to-one comparison between the two sets of maps
is not possible  since different smoothing procedures were
used and since our residuals are computed at the PSC$z$ predicted distances
whereas the ones of D98 are computed at the redshift space positions.


\subsection{Residual Correlation Function}
\label{sec:corrfunc}

A more quantitative assessment of the goodness of fit can be obtained by 
computing the correlation function of the unsmoothed magnitude
residuals:
\begin{equation} 
\psi(\tau)={{1}\over{N(\tau)}} \sum_{i<j}\delta_{m,i}\delta_{m,j},
\label{corre}
\end{equation}
where $N(\tau)$ is the number of galaxy pairs with predicted
separation 
$d_{ij} \le \tau \pm 100 \kms$. This correlation function
applies to normalized magnitude residuals, i.e., does not depend on 
$\sigma_v$ and $\sigma_{TF}$ and is only sensitive to genuine correlation
among residuals. 

\begin{figure}
\vspace{11.0truecm}
{\includegraphics{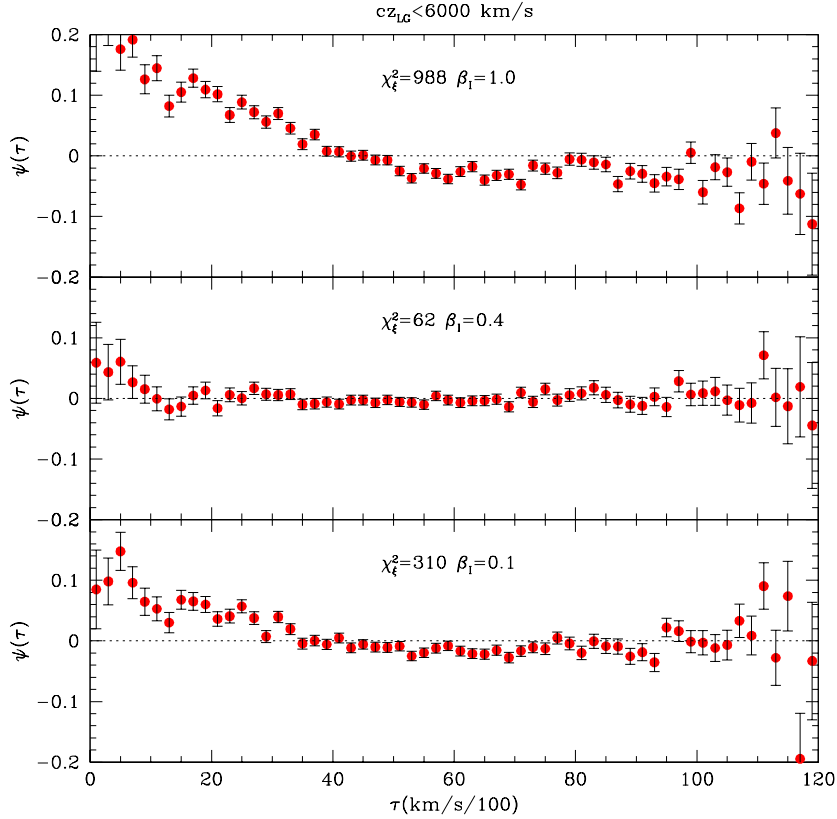}}
\caption{\capt The correlation function of magnitude residuals 
plotted for $\beta_I=0.1$ (lower plot), $\beta_I=0.4$ (middle plot)
and $\beta_I=1.0$ (upper plot). The results refer to the full SFI
sample. }
\label{fig:corrbeta}
\end{figure}

In fig.~\ref{fig:corrbeta} we show the correlation function of all
SFI galaxies with $cz_{LG}<6000$ \kms for the same three values of
$\beta_I$
used to produce the velocity residuals shown in figs.~\ref{fig:vmap04},  
\ref{fig:vmap01} and ~\ref{fig:vmap1}. The error bars represent Poisson
errors $N(\tau)^{-0.5}$.  In the $\beta_I=0.4$ model, the correlation
function appears to be consistent with zero almost everywhere apart
from some positive correlation at separations smaller than 500 \kmsn,
i.e. of the order of the smoothing scale of the model velocity
field. A significant excess correlation on small and large scales is
detected for $\beta_I=0.1$ and, to an even greater degree, for
$\beta_I=1.0$.

\begin{figure}
\vspace{11.0truecm}
{\includegraphics{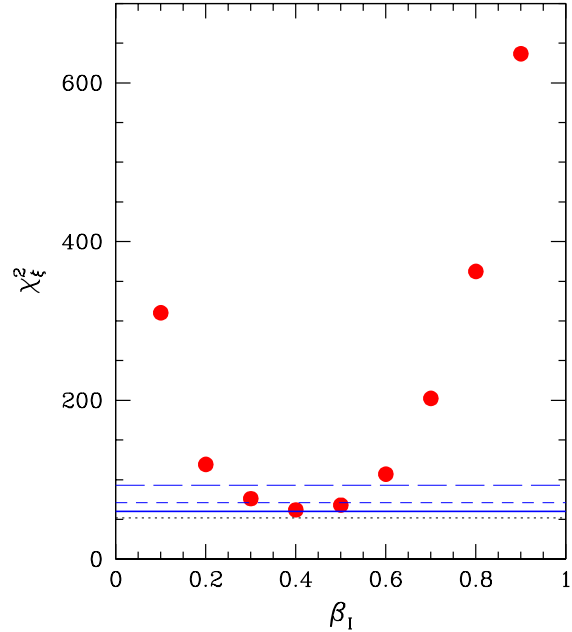}}
\caption{\capt The statistic $\chi^2_{\xi}$ plotted for various values of 
$\beta_I$. The heavy line shows the expectation value for a $\chi^2$
statistic with the same number of degrees of freedom. 
The dashed and long--dashed lines show 1$\sigma$ and 3$\sigma$
deviations
from the expectation value.
The dotted line shows
the expectation value of  $\chi^2_{\xi}$ corrected according to V1.
}
\label{fig:chi}
\end{figure}

To quantify the goodness of the fits we compute the quantity
\begin{equation}
\chi^2_{\xi}=\sum_{i=1}^{N_{bins}}
{{\xi^2(\tau_i)}\over{N(\tau_i)}},
\label{chi}
\end{equation}
where $\xi(\tau)=N(\tau)\psi(\tau)$ and $N_{bins}$ is the number of
independent distance bins in which $\psi(\tau)$ is computed.  V1 have
shown that if the residuals are indeed uncorrelated on a scale $\tau$
then $\xi(\tau)$ is a Gaussian random variable with zero mean and
variance $N(\tau)$. Under this approximation, the $\chi^2_{\xi}$ 
statistic is distributed as $\chi^2$ with number of 
degrees of freedom equal to the
number of independent distance bins.  Any correlation among residuals
will result in a larger $\chi^2_{\xi}$. Extensive tests with mock
catalogues performed by V1 revealed that $\chi^2_{\xi}$ indeed
has properties similar to a $\chi^2$ statistic, with the same
variance, 
but with a mean of $\sim 0.87$ per degree of freedom rather than
unity. We have computed the quantity $\chi^2_{\xi}$ for 10 values of
$\beta_I$ ranging from 0.1 to 1.0. The results are shown in
fig~\ref{fig:chi}. The continuous, heavy line shows the expectation
value for a $\chi^2$ statistic with $N_{bins}=60$ degrees of freedom, 
while the dashed and long--dashed lines show, 
respectively, the 1$\sigma$ and 3$\sigma$ 
deviations from that value.  The dotted line represents the
expectation value of $\chi^2_{\xi}$ according to the V1 correction, i.e
assuming a number of degrees of freedom of $0.87\times 60 = 52.2$.

Despite its limited discriminatory power, the 
$\chi^2_{\xi}$ statistic 
clearly indicates that PSC$z$ velocity models with $0.3 \le 
\beta_I \le 0.5$, and therefore also 
our best model according to the VELMOD analysis, $\beta_I = 0.42 \pm 0.04$, 
provide an acceptable fit to the SFI velocities. 
All other models with smaller or larger values of  $\beta_I$
can be ruled out at a level $>3$$\sigma$.

\begin{figure}
\vspace{11.0truecm}
{\includegraphics{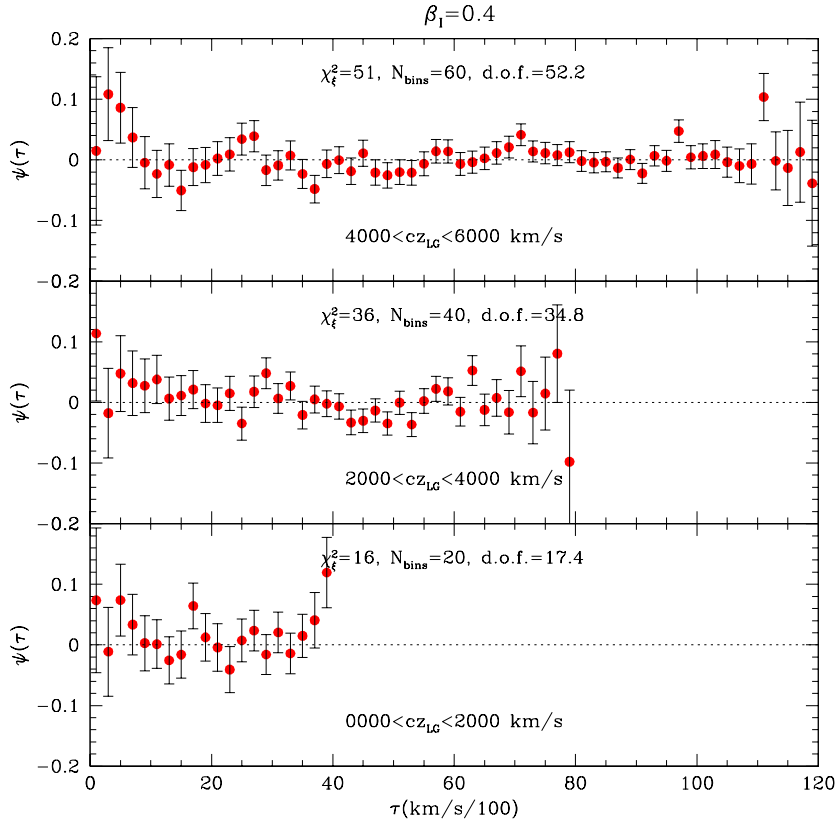}}
\caption{\capt Residual correlation function in three redshift shells for
the model
$\beta_I=0.4$. 
The redshift ranges, values of $\chi^2_{\xi}$, number of bins 
and, in parenthesis, the number of degrees of freedom corrected according
to V1 are
also shown in the plots
}
\label{fig:corrz}
\end{figure}

We are now in a position to address the question of whether the
PSC$z$ model velocity field provides an acceptable fit throughout the
whole SFI sample, particularly in the external regions where
$\sigma_v$ is large.  We do that by computing the residual
correlation function in the same three redshift shells previously
used.  The results are displayed in fig.~\ref{fig:corrz} which shows
the correlation functions, $\psi(\tau)$, for the three subsamples.

The value of $\psi(\tau)$ appears to be consistent with the null
hypothesis of no correlation  in nearly all the distance bins in each
of the three redshift shells, with the exception of an excess
correlation at small separation in the outermost shell.  The value of
the statistic $\chi^2_{\xi}$ in each shell is indicated in the
plots. It is always smaller than the number of bins used in each
shells and comparable to the expectation value for $\chi^2_{\xi}$
computed using the V1 correction.
These results show that,
despite the large value of $\sigma_v$ obtained from
the VELMOD analysis, the residuals of the fit are not correlated,
i.e. the differences in magnitude between model and data
are randomly distributed.

\section{Discussion}
\label{sec:disc}

The results of our VELMOD analysis generally agree with those of
independent analyses, with the exception of the velocity noise,
$\sigma_v$. We find a significantly larger value than that found by V1
and V2 who compared the Mark~III velocities to the IRAS 1.2Jy model
velocity field.  A VELMOD analysis of SFI subsamples limited in
redshift revealed the existence of two regions.  An inner sphere of
radius 4000 \kms in which $\sigma_v\simeq 200$ \kmsn, a value which is
believed to reflect the cumulative effect of velocity noise and errors
in the model predictions (as corroborated by the results of the
analysis of the A82 and MAT samples), and an external region in which
$\sigma_v \sim 500 \kms$.

If these large values of $\sigma_v$ were caused by systematic
deviations of the SFI velocity field from the model predictions, then
the residuals should correlate on scales larger than the smoothing
length. However, this is not the case (see fig.~10).  In fact, the
residuals turn out to be quite small and do not show any significant
spatial correlation.  This is true for all redshift shells, including
the outermost one where $\sigma_v$ is large.  Moreover, a quantitative
analysis based on a $\chi^2$ statistic shows that the PSC$z$ velocity
model with $0.3<\beta_I<0.5$ does provide a satisfactory fit to the
SFI data without the need to introduce any external fields as in the
V1 and V2 VELMOD analyses.  This can be interpreted as implying that
discrepancies between the model and the data do exist, but not in the
form of peculiar motions coherent on scales larger than $\sim 300$
\kmsn. Rather, they are randomly distributed and thus properly
quantified by means of the total variance, to which both $\sigma_v$ and
$\sigma_{TF}$ contribute.  It is worth noticing that the increase in 
$\sigma_v$ occurs at the distance at which large structures like the
Perseus-Pisces supercluster and the Great Attractor appear in the
sample.  Non-linear effects may be strongest in these structures,
leading to a mismatch on small scales between measured and
predicted velocity fields which could contribute to the increase in
$\sigma_v$. 
To investigate whether these structures are indeed responsible
for the increase in $\sigma_v$ we have ran two VELMOD analyses
in which we have excluded either SFI galaxies in the Perseus-Pisces
supercluster
(i.e. with $120^{\circ} \le l \le 170^{\circ}$,
$-35^{\circ}\le b \le 5^{\circ}$ and $4000 \kms \le cz_{LG}\ le 6000 \kms$
or in the Hydra Centaurus region
(i.e. with $260^{\circ}\le l \le 330^{\circ}$,
$-10^{\circ}\le b \le 40^{\circ}$ and $2000 \kms \le cz_{LG} \le 6000
\kms$).
A third analysis has been performed after excluding galaxies in both
structures.
The results turned out to be very similar to those shown in figure 4 for
the
full sample case (the differences are of the order of 20 \kmsn).
The abrupt increase of $\sigma_v$ at
$cz_{LG} \ge 4000 \kms$ is present in all the cases explored,
showing that  neither the Perseus-Pisces supercluster nor the Great
Attractor contribute appreciably  to the variation of $\sigma_v$.

A second possibility is that the increase in $\sigma_v$ with
redshift is caused by a comparable increase in the errors of the
velocity model. Indeed, B99 quantified such effect (see their eqn.~18)
which, however, is far too small to explain the 
observed  increase in $\sigma_v$.

The final possibility is that the TF scatter decreases significantly
with distance, which seems reasonable, since the scatter tends to
decrease for high-linewidth galaxies, and these are seen
preferentially at large distances.  Because of the covariance between
$\sigma_v$ and $\sigma_{TF}$ (see V1), this could lead to an
overestimation of $\sigma_v$ and an underestimation of $\sigma_{TF}$.
In this context, it is interesting to note that a power spectrum
analysis of the same SFI dataset, like the one performed by Freudling
et al. (1999), also returns different results if computed separately
for the closer and more distant halves of the sample.  However, such
analysis reveals that it is the inner part of the sample which has an
apparently enhanced $\sigma_v$.  Given these conflicting results, we
consider it unlikely that the apparent increase in $\sigma_v$ reflects
a true property of the velocity field.  In any case, it is worth
stressing that the increase of $\sigma_v$ with distance does not
affect the estimate of $\beta_I$ which, as shown in fig.~4 and
table~3, does not show significant variations.

\section{Summary and Conclusions}
\label{sec:conc}

We have implemented, tested and used the VELMOD technique (V1, V2) to
compare the model velocity field obtained from the spatial
distribution of PSC$z$ galaxies with the velocities of field spiral
galaxies within $6000$ \kms in the SFI catalogue.  This comparison
allowed us to estimate the value of the $\beta_I$ parameter, the
amplitude of the small-scale velocity noise, $\sigma_v$, and to
calibrate the TF relation of SFI galaxies.

VELMOD returns an estimate of $\beta_I$ which is very robust to
various systematic and random errors that may enter the analysis at
various stages.  For  the full sample, we obtain $\beta_I= 0.42\pm
0.04$, while a value of $\beta_I= 0.45\pm 0.05$ 
is obtained  when a more conservative cut at 
$4000$  \kms is applied to the SFI sample.

The slope of our TF relation is in good agreement with that obtained
from the calibration of the TF relation by G97 using SCI galaxies.
However, we find a significantly larger TF scatter ($\sigma_{TF} =
0.44$) than the average scatter found by G97 ($\sigma_{TF} = 0.36$),
and a significantly larger zero point ($A_{TF}=-5.89$ {\em vs}
$A_{TF}=-6.09$).  We thus conclude that the scatter in the TF relation
for field galaxies is larger and the zero point is smaller than the
correspoding values for galaxies in clusters, confirming the results
of D98 and Freudling \etal (1999).  Our analysis also confirms that
the TF scatter decreases with increasing galaxy rotation velocity, but
this does not affect our estimates of $\beta_I$.  We have also found
that the velocity noise, $\sigma_v$, increases with distance up to
$\sim 500$ \kms and is significantly larger than the values obtained
by V1 and V2. Although the meaning of such a large $\sigma_v$ is not
clear, it is reassuring that it does not affect the general result of
the VELMOD analysis and, in particular, the value of $\beta_I$. 

In summary, we have found that the PSC$z$ velocity model provides a
satisfactory fit to the velocities of SFI galaxies when a value of
$\beta_I=0.42 \pm 0.04$ is used. This value is in good agreement with
most of the recent v-v analyses that use the IRAS gravity field,
inferred 
either from the 1.2Jy or the PSC$z$ surveys, irrespective of the type of
distance indicator used. Indeed, values of $\beta_I$ in the range 0.4
- 0.6 are found using the TF relation for spiral galaxies (Davis
\etal  1996, V1, V2, D98), the $D_n - \sigma$ relation for
early type galaxies (Nusser \etal 2000), Type Ia Supernovae (Riess
\etal 1997) and the ``surface brightness fluctuations'' method in
nearby galaxies (Blakeslee \etal 2000).  Our estimate of $\beta_I$ is
also consistent with the results of v-v analyses which use a model
gravity field derived from optical galaxies in the ORS survey
(Santiago \etal 1995, 1996), once the stronger clustering of optical
galaxies is taken into account.

These values of $\beta_I$ from v-v analyses are significantly smaller
than those found using d-d methods (e.g. Sigad \etal 1998), even when
both are applied to identical datasets.  Further support for high
values of $\beta$ comes from power spectrum analyses of various
velocity catalogues (Zaroubi \etal 1997, Freudling \etal 1999, Zaroubi
\etal 2000), although in this case it is possible that accounting for
nonlinear effects may help reduce the discrepancy (Silberman \etal
2000).
Since v-v methods generally rely on the deviations of a set of
observed velocities from a set of model velocities, differences
between the two sets do not contribute to $\beta$ but to the random
errors, whereas this is not the case in power spectrum analyses.

A different explanation for the discrepancy between v-v and d-d
methods is that the v-v analyses are affected by non-linear effects.
All v-v methods implemented so far assume linear biasing and most of
them also use linear gravitational instability theory.  Recently new
methods have been introduced to measure the nonlinearity in the bias
relation in redshift surveys (e.g. Matarrese, Verde \& Heavens 1997,
Szapudi 1998, Sigad, Branchini \& Dekel 2000, Feldman \etal 2000) and
some of these have been applied to the PSC$z$ survey.  For example,
Branchini \etal (in prep.) have applied the technique of Sigad \etal
(2000) to the PSC$z$ sample. Their preliminary results indicate that
deviations from linear biasing are small and manifest themselves
mainly as a anti-bias in low density regions. The amplitude of these
effects can only lead to a modest increase of $\beta_I$, of the order
of $\sim 10\%$.  Nonlinear motions might also change the value of
$\beta_I$.  However, Shaya, Peebles \& Tully (1995) used a fully
nonlinear model of the velocity field and still recovered a low value of
$\beta$.  Similarly, V1 performed a nonlinear VELMOD analysis in an
attempt to break the degeneracy between $\Omega_m$ and $b_I$ and found
indirect evidence that nonlinear motions are already accounted for in
the model velocity field when this is predicted from a smooth 
density field.  These results suggest that a treatment of nonlinear
effects and the inclusion of nonlinear biasing prescriptions in v-v
analyses will not change the inferred value of $\beta$ significantly.
Thus, it seems unlikely that the disagreement between the two methods
will be eliminated purely by improving the existing v-v techniques.

\section*{Acknowledgments}

It is a pleasure to acknowledge 
Avishai Dekel, Adi Nusser and Saleem Zaroubi 
for many useful discussions. A special 
thanks goes to Michael Strauss for his 
very useful suggestions and for providing 
some of the Mark III subsamples used
in  the original VELMOD analysis. 
For these same reasons we would have liked to thank Jeff Willick
who died in tragic circumstances.
EB thanks ESO for its hospitality,
where part of this work was done.
This work has been partially supported by US NSF grants AST--9528860
to MPH, AST--9617069 to RG and AST--9900695 to RG and MPH and by a UK
PPARC rolling grant to CSF. 
IZ was supported by the DOE and the NASA grant NAG 5-7092 at Fermilab.

\newpage


\begin{table}
\centering
\caption[]{VELMOD analysis of mock catalogues. 
The top row lists the value of the true parameters.
The second row lists 
the mean value, the error on the mean 
and, in parenthesis, the typical error in a single realization
from VELMOD.
Column 1: TF zero point
; column 2: TF slope; column 3: TF dispersion;
column 4: 1D velocity dispersion (in \kms); column 5: value of $\beta_I$.}
\tabcolsep 2pt
\begin{tabular}{ccccc} \\  \hline
$A_{TF}$ & $b_{TF}$&
$\sigma_{TF}$ & $\sigma_{v}$ & $\beta_I$ \\ \hline
-6.1 & 7.33& $0.36$ & $150$ & $1.0$ \\ 
$-6.096\pm0.005 (0.02)$& $7.31\pm0.04 (0.16)$ 
& $0.36\pm0.002 (0.01)$ & $150\pm7 (33)$ & $1.01\pm0.009 (0.04)$ \\ 
\hline 
\end{tabular}
\label{tab:mock}
\end{table}

\begin{table}
\centering
\caption[]{VELMOD analysis of Mark III subcatalogues. The 
corresponding 
results of the V1 and V2 analyses are quoted in parenthesis.
Column 1: Subcatalogue;
column 2: TF zero point; column 3: TF slope; column 4: TF dispersion;
column 5: 1D velocity dispersion (in \kms); column 6: $\beta_I$
value and its 1-$\sigma$ error}
\tabcolsep 2pt
\begin{tabular}{cccccc} \\  \hline
Catalogue & $A_{TF}$ & $b_{TF}$&
$\sigma_{TF}$ & $\sigma_{v}$ & $\beta_I$ \\ \hline
A82 & -5.91 (-5.96) &  10.43 (10.36) 
& 0.468 (0.464) & 159 (125) & $0.50\pm0.07 \ (0.49\pm0.08)$ \\ 
MAT & -5.84 (-5.75) &  7.06 (7.12) 
& 0.451 (0.453) & 180 (125) & $0.42\pm0.10 \ (0.50\pm0.11)$ \\ 
MAT2 & -5.83 (-5.80) &  6.96 (7.16) 
& 0.433 (0.430) & 220 (130) & $0.44\pm0.07 \ (0.52\pm0.05)$ \\ 
\hline
\end{tabular}
\label{tab:mk3}
\end{table}

\begin{table}
\centering
\caption[]{VELMOD analysis of various SFI subsamples.
Column 1: redshift range (in \kms) and additional cuts; column 2: number
of galaxies
column 3: TF zero point; column 4: TF slope; column 5: TF dispersion;
column 6: 1D velocity dispersion (in \kms); column 7: $\beta_{min}$ and
its 1-$\sigma$ error}
\tabcolsep 2pt
\begin{tabular}{ccccccc} \\  \hline
Sample &  N$_{\rm gal.}$ & $A_{TF}$ & $b_{TF}$ &
$\sigma_{TF}$ & $\sigma_{v}$ & $\beta_I$ \\ \hline
$cz_{LG}< 6000$ & 989 &  -5.89 &   7.19
& 0.439 & 250 & $0.42\pm0.04$\\ 
$cz_{LG}< 6000, \ \eta>-0.1$ & 640 &  -5.89 &   6.74
& 0.382 & 230 & $0.43\pm0.04$\\ 
$cz_{LG}< 6000, \ \eta<-0.1$ & 349 &  -5.87 &   7.04
& 0.498 & 248 & $0.39\pm0.04$\\ 
$cz_{LG}< 4000$ & 493 &  -5.86 &   7.32
& 0.415 & 216 & $0.45\pm0.05$\\  
$cz_{LG}< 2000$ & 158 &  -5.73 &   6.67
& 0.454 & 206 & $0.42\pm0.09$\\
$2000 \le cz_{LG}< 4000$ & 335 &  -5.88 &   7.41
& 0.410 & 196 & $0.44\pm0.07$\\
$4000 \le cz_{LG}< 6000$ & 496 &  -5.85 &   7.54
& 0.420 & 535 & $0.40\pm0.09$\\
\hline
\end{tabular}
\label{tab:res}
\end{table}

\end{document}